\documentclass[10pt,a4paper,twocolumn]{article}

\usepackage{authblk}
\usepackage{graphicx}
\usepackage[explicit]{titlesec}
\usepackage[labelfont=bf,labelsep=endash,font=footnotesize]{caption}
\usepackage{tabu}
\usepackage{xcolor}
\usepackage[hidelinks, bookmarks=false]{hyperref}
\usepackage[hang]{footmisc}
\usepackage[normalem]{ulem}
\usepackage[top=2.5cm, bottom=2.8cm, left=1.5cm, right=1.5cm]{geometry}
\usepackage{abstract}
\usepackage{mathtools}
\usepackage{balance}
\usepackage{cite} 
\usepackage{wrapfig} 
\usepackage{array} 
\usepackage{pifont} 

\makeatletter
\renewcommand\AB@affilsepx{ \protect\Affilfont}
\makeatother

\providecommand{\keywords}[1]{\textbf{Keywords}\ \ \textendash\ \   #1}

\titleformat{\section}{\large\bfseries}{\thesection.}{1em}{\MakeUppercase{#1}}
\titlespacing*{\section}{0pt}{12pt}{6pt}

\titleformat{\subsection}{\large}{\thesubsection}{1em}{#1}
\titlespacing*{\subsection}{0pt}{12pt}{6pt}

\titleformat{\subsubsection}{\large\itshape}{\thesubsubsection}{1em}{#1}
\titlespacing*{\subsubsection}{0pt}{12pt}{6pt}

\newcommand{\ITUurl}[1]{\textcolor{blue}{\urlstyle{same}\url{#1}}}

\newcommand{\ITUpar}{\vspace{8pt}\par}

\renewenvironment{abstract}
               {\list{}{
               \setlength{\rightmargin}{0mm}
               \setlength{\leftmargin}{0mm}
               \vspace{-0.25in}
                \item[\textit{\textbf{\hspace{22pt}Abstract  }}  \textendash]\relax}}
               {\endlist}

\def\starttable{\vspace{6pt}\begin{table}[ht]\center}
\def\startfigure{\vspace{6pt}\begin{figure}[ht]\center}

\makeatletter
\def\tagform@#1{\maketag@@@{\ignorespaces#1\unskip\@@italiccorr}}
\makeatother

\title{\large{\textbf{\uppercase{6G VISION: AN ULTRA-FLEXIBLE PERSPECTIVE}}}}

\author[1]{\normalsize{Ahmet Yazar}}
\author[1]{\normalsize{Seda Do\u{g}an Tusha}}
\author[1,2]{\normalsize{Huseyin Arslan}}

\affil[1]{\normalsize{Department of Electrical and Electronics Engineering, Istanbul Medipol University, Istanbul, 34810, Turkey}}
\affil[2]{\normalsize{Department of Electrical Engineering, University of South Florida, Tampa, FL, 33620, USA}}
\affil[ ]{\protect\\}
\affil[ ]{\protect\\  \normalsize{Corresponding author: Ahmet Yazar (ayazar@medipol.edu.tr)}}
\affil[ ]{\protect\\}
\affil[ ]{\protect\\  \normalsize{This work was published in ITU Journal on Future and Evolving Technologies - Volume 2020. http://handle.itu.int/11.1002/pub/8173e1cc-en}}

\date{\vspace{-12pt}\endgraf\rule{\textwidth}{1pt}}

\begin{document}

\cleardoublepage

\twocolumn[

\begin{@twocolumnfalse}

\maketitle

\begin{abstract}
\textit{The upcoming sixth generation (6G) communications systems are expected to support an unprecedented variety of applications, pervading every aspect of human life. It is clearly not possible to fulf\kern 0.06667emill the service requirements without actualizing a plethora of f\kern 0.06667emlexible options pertaining to the key enabler technologies themselves. At that point, this work presents an overview of the potential 6G key enablers from the f\kern 0.06667emlexibility perspective, categorizes them, and provides a general framework to incorporate them in the future networks. Furthermore, the role of artif\kern 0.06667emicial intelligence and integrated sensing and communications as key enablers of the presented framework is also discussed.}
\end{abstract}

\ITUpar

\keywords{6G, adaptive, artificial intelligence, cognitive radio, dynamic, flexibility, sensing.}

\ITUpar
\ITUpar

\end{@twocolumnfalse}
]

\section{Introduction} 
\label{sec:sec1}

Following the successful standardization of the Fifth Generation (5G) networks worldwide, academia and industry have started to turn their attention to the next generation of wireless communications networks \cite{9145564}. At present, there are more than 100 research papers regarding the Sixth Generation (6G) of wireless communications. There is no doubt that new visions and perspectives will continue to be developed in the coming years. However, despite all these efforts, current literature lacks gathering the distinctive features of 6G under a single broad umbrella.\\

\begin{figure}
	\centering
	\includegraphics[width=7.5cm]{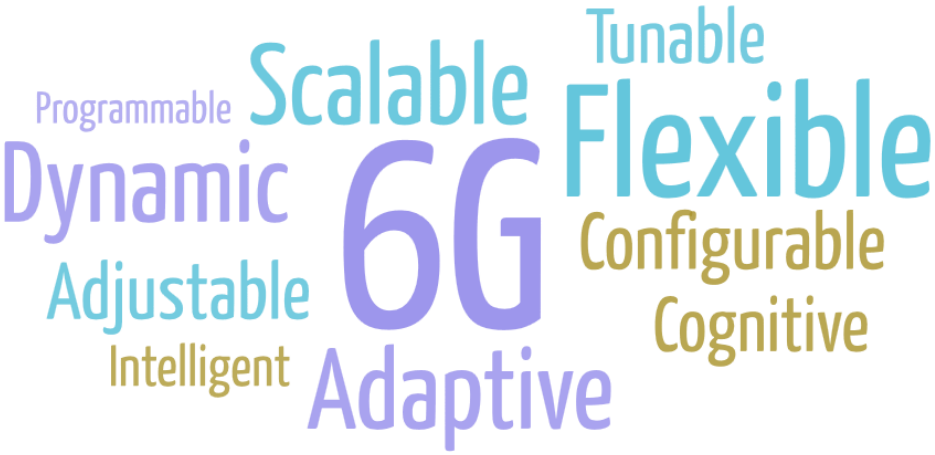}
	\caption{Flexibility terms.}
	\label{fig:s1a}
\end{figure}

\begin{figure*}[thp!]
    \centering
	\includegraphics[width=17.6cm]{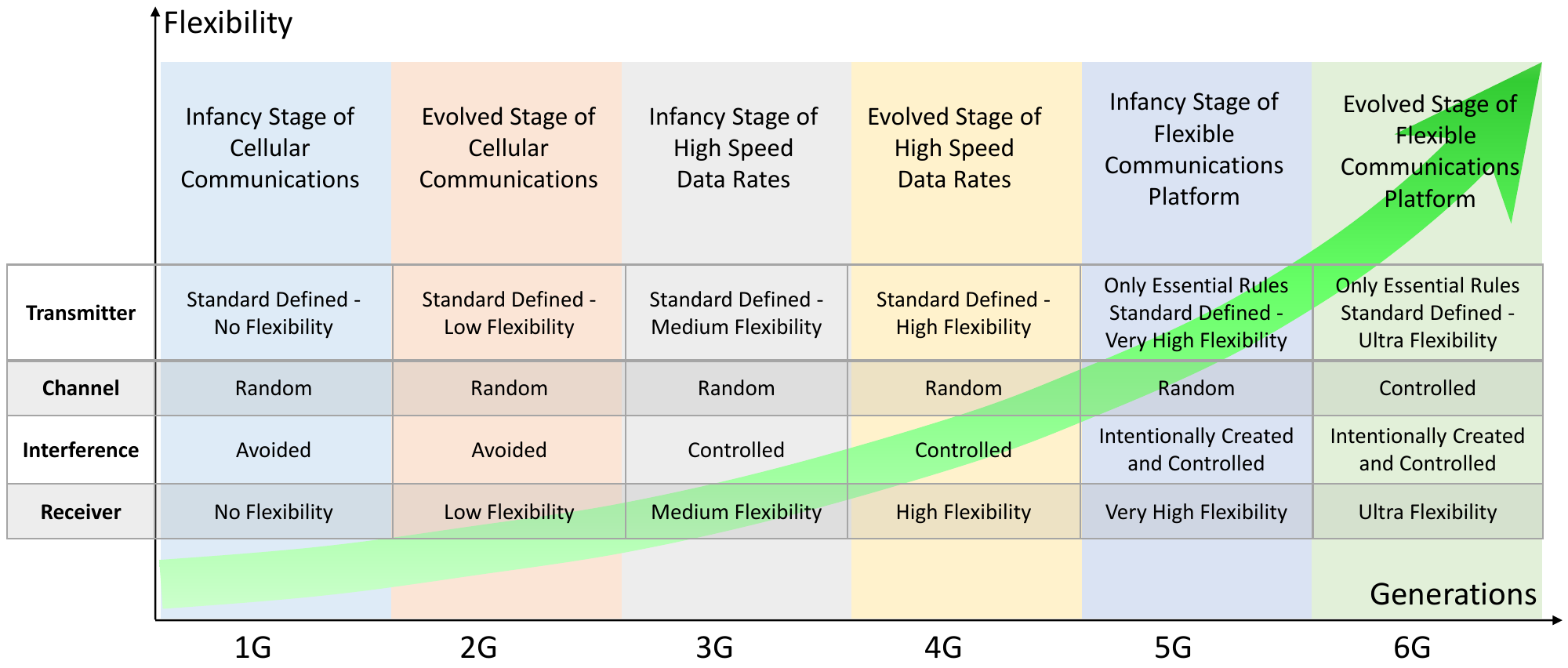}
	\caption{Flexibility analysis of the previous generations and 6G communications.}
	\label{fig:s1c}
\end{figure*}

The evolution of cellular communications through different generations from the Radio Access Technology (RAT) perspective is shown in Table~\ref{fig:s1b}. The number of capabilities for newer cellular generations increases as a result of the need to meet diversified requirements. Flexibility\footnote{The other terms used interchangeably for flexibility are shown in Fig.~\ref{fig:s1a}.}, where it is defined as the capability of making suitable choices out of available options depending on the internal and external changes, of the communications systems eventually evolves with an increasing number of new options. In this context, Fig.~\ref{fig:s1c} provides a concise flexibility analysis for different generations of cellular communications considering the features in Table~\ref{fig:s1b}.\\

The Second Generation (2G) systems have paved the way for flexibility in communications systems by means of multiple frequency reuse options, adaptive equalization, and dynamic channel allocation. The Third Generation (3G) and the Fourth Generation (4G) cellular systems have incorporated voice communications with data communications. Additionally, Code Division Multiplexing (CDMA) and Orthogonal Frequency Division Multiplexing (OFDM) have provided flexibility in terms of multiplexing, rate adaptation and interference management via the exploitation of different spreading factors and the multidimensional resource utilization, respectively.\\

\begin{table*}[ht!]
	\begin{center}
		\begin{tabular}{c}
			\raisebox{-\totalheight}{\includegraphics[width=17.6cm]{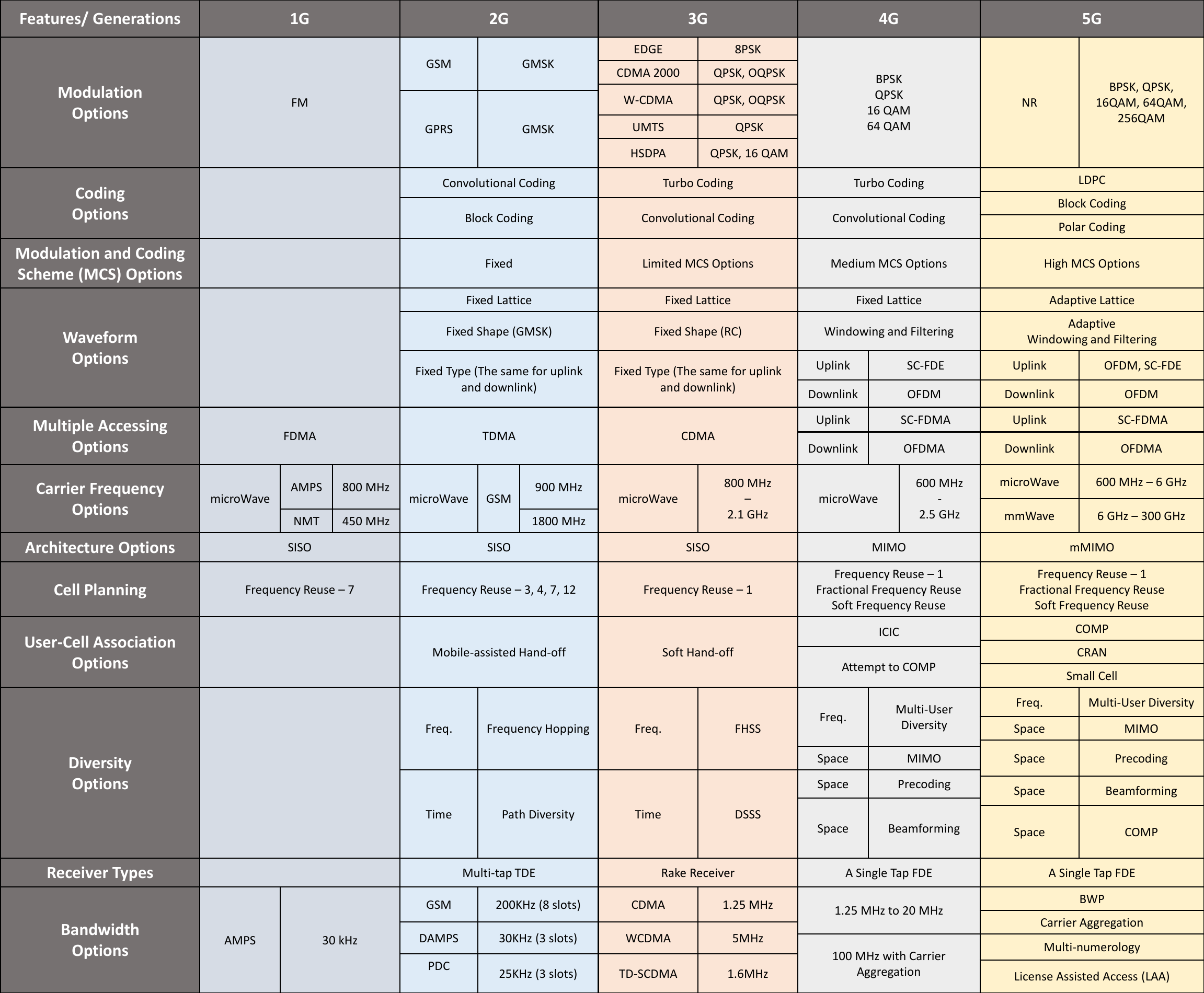}}
		\end{tabular}
		\caption{Increasing number of features for cellular generations.}
		\label{fig:s1b}
	\end{center}
\end{table*}

The introduction of various services with rich requirement sets under 5G has revealed the need for a flexible network that can simultaneously meet diverse requirements. 5G has given a start for  flexible wireless communications by the accommodation of different technologies. To exemplify, the coexistence of multi-numerology in a single frame has been adopted during standardization meetings. In a given network, achieving flexibility is mainly dependent on three capabilities \textit{1) awareness, 2) availability of a rich set of technology options,}  and \textit{3) adaptation \& optimization}. On this basis, although the flexibility perspective has been broadened in 5G systems with respect to previous generations, the existing technology options are not enough to reach all the goals of 5G networks. Additionally, it is expected that 6G networks  will put further pressure on service providers due to emerging  applications and use cases corresponding to new sets of requirements. Therefore, 6G systems need to extend the current flexibility by \textit{(1)} exploring the awareness for the different aspects of the whole communications network and environment using  different sensing mechanisms including Artificial Intelligence (AI), \textit{(2)} enriching technology options, and \textit{(3)} providing optimum utilization of available options considering the awareness with practical sensing capabilities.\\

The work scopes of 6G publications in the literature are summarized in Table~\ref{tab:new}. A majority of the current 6G related studies attempt to identify the future applications and their key requirements \cite{9145564, 9178307, 8760275, 8766143, 8820755, 8869705, 9023459, 9040264, 9124725, dang2020, alsharif2020, 9170653, 9205980, 8782879, 8959607, 9083880, yuan2020, 9163104, 9224777, 9225680, 8412482, 9083759, 8941882, 9086145, 9184022, 9210383, 8808168, 8681450, 8760269, 9042251, 9083770, 9083784, 9146540, 9151546, 9205981, 9206115, 9237116, S2352864820300237}. Moreover, potential service types and application groups for 6G are analyzed in \cite{9145564, 8760275, 8766143, 8820755, 8869705, 9023459, 9040264, 9124725, dang2020, alsharif2020, 9170653, 9205980, 8941882, 9086145, 9184022, 9210383, 8808168} together with the prospective key requirements of 6G networks. Several works are focused on the key enabler technologies and concepts under 6G studies in detail \cite{9145564, 9178307} or in general \cite{8760275, 8766143, 8820755, 8869705, 9023459, 9040264, 9124725, dang2020, alsharif2020, 9170653, 8782879, 8959607, 9083880, yuan2020, 9163104, 9224777, 9225680}. Furthermore, specific technologies and concepts are also being pushed for 6G as described in \cite{8941882, 9086145, 9184022, 9210383, 8808168, 8681450, 8760269, 9042251, 9083770, 9083784, 9146540, 9151546, 9205981, 9206115, 9237116, S2352864820300237, 8732419, 8760401, 8798636, 8922617, 8926369, 8981888, 9049790, 9055054, 9061001, 9083780, 9083786, 9083794, 9083805, 9136592, 9174846, 9183763, 9193934, 9201361, 9205279, 9205314, 9205906, 9208801, 9217361, 9221119, 9222142, 9222169, 9229054, 9237460, 9239396, 9241414, 9247128, 9247315, 9247451, 9247527, 9083851, 9040202, 9122618, 9200631, S2352864819304274, S2352864820300249, S2352864820302418, S2352864820302431, S2352864820301802, S235286482030242X}. These studies are revisited in the next two sections, however, it is seen that the flexibility perspective of the key enablers is not considered as a distinguishing feature for 6G systems in the literature.\\

In light of the aforementioned discussions, 6G networks require the redesign of cellular communications to provide extreme flexibility in all of its building blocks. Correspondingly, this paper elaborates the example flexibility aspects of potential 6G key enablers and provides a unique categorization of the related technologies and concepts. Moreover, a novel framework is proposed to gather the said enablers under an umbrella of a single ultra-flexible framework for 6G.\\

The rest of the paper is organized as follows: Section~\ref{sec:s2} gives a brief overview for the initial forecasts on 6G to explain the background for the necessity of a flexible perspective without examining all potential applications, requirements, and service types. Flexibility discussions on the potential 6G key enablers are provided under a unique categorization rather than giving different details of these enablers in Section~\ref{sec:s3}. The scope of Section~\ref{sec:s3} is limited to example flexibility aspects for the potential 6G key enablers. Next, a framework is proposed to increase interoperability of the 6G enablers in Section~\ref{sec:s4}. Finally, conclusions are drawn with several open issues in Section~\ref{sec:s5}.

\section{A Brief Overview: Forecasts on 6G}
\label{sec:s2}

\begin{figure}[b!]
\centering
\includegraphics[width=8cm]{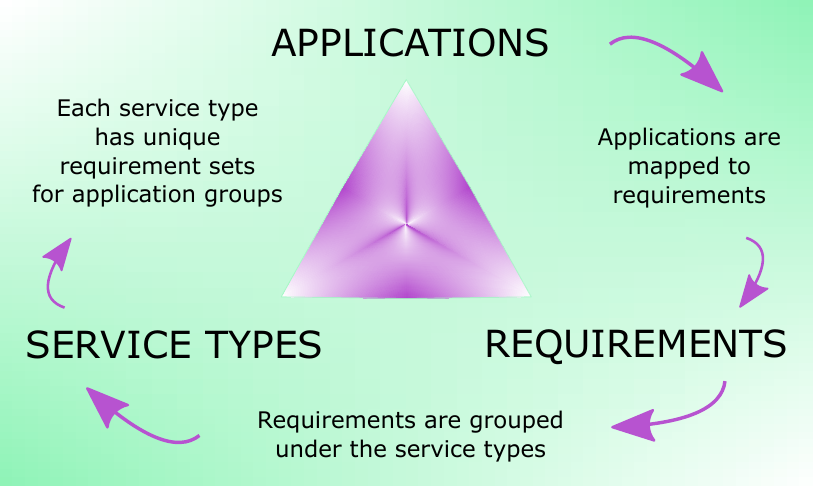}
\caption{A basic relationship between the applications, requirements, and service types.}
\label{fig:s2a}
\end{figure}

Identification of the future applications, requirements and possible service types is one of the primary objectives of the initial 6G research studies. Fig.~\ref{fig:s2a} illustrates the basic relationship between these components. Mapping the potential future applications to the several requirements with different priorities is accepted as a first step in general. Next, these requirements are grouped under the service types in a reasonable manner. At the final stage, service types have unique requirement sets for the related application groups. In 5G systems, applications are considered under three service types including enhanced Mobile BroadBand (eMBB), Ultra-Reliable and Low-Latency Communications (URLLC), and massive Machine-Type Communications (mMTC) \cite{itu2015}. Among these, eMBB applications prioritize high throughput, capacity and spectral efficiency; mMTC prioritizes energy efficiency and massive connectivity while URLLC requires high reliability and low latency. For 6G systems, some of the initial studies inherently analyze the relations between the future applications and prioritized requirements to propose candidate service types \cite{9145564, 9178307, 8760275, 8766143, 8820755, 8869705, 9023459, 9040264, 9124725, dang2020, alsharif2020, 9170653, 9205980, 8782879, 8959607, 9083880, yuan2020, 9163104, 9224777, 9225680, 8412482, 9083759, 8941882, 9086145, 9184022, 9210383, 8808168, 8681450, 8760269, 9042251, 9083770, 9083784, 9146540, 9151546, 9205981, 9206115, 9237116}.\\

\begin{table*}[ht!]
  \begin{center}
    \begin{tabular}{c}
      \raisebox{-\totalheight}{\includegraphics[width=17.6cm]{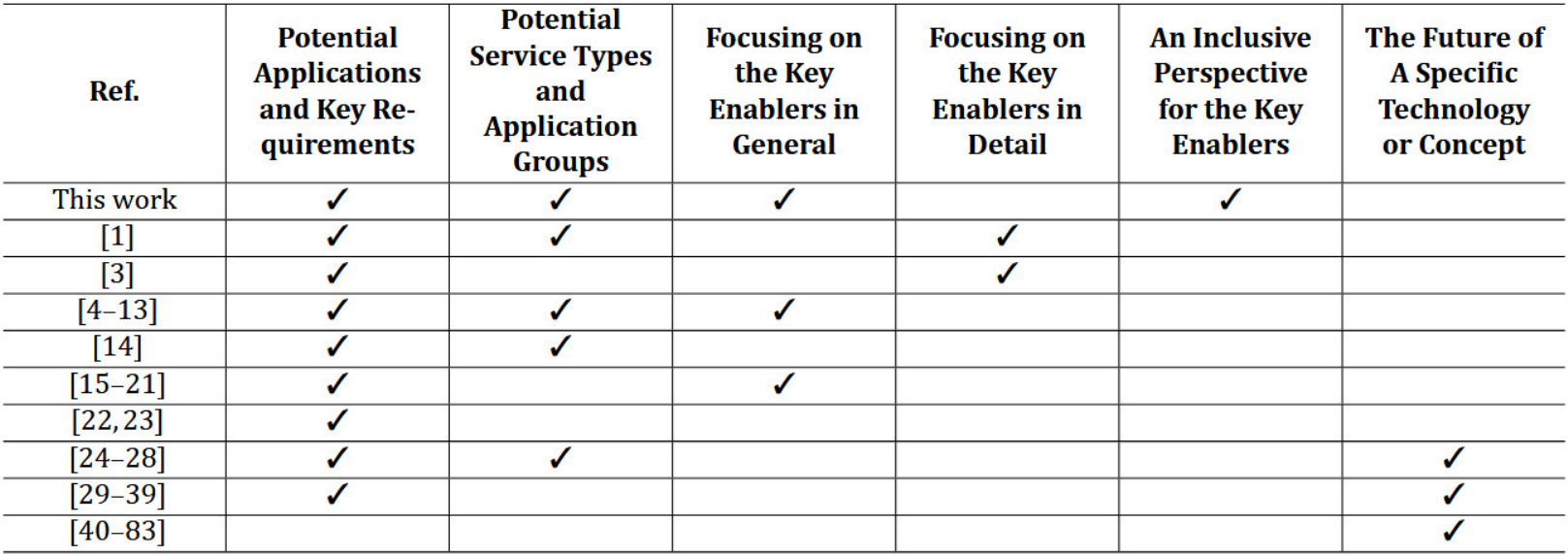}}
    \end{tabular}
    \caption{Scopes of 6G publications in the literature.}
    \label{tab:new}
  \end{center}
\end{table*}

The following list exemplifies potential 6G applications: drone and Unmanned Aerial Vehicle (UAV) networks, drone taxi, fully automated Vehicle-to-Everything (V2X), remote surgery, health monitoring, e-health, fully sensory Virtual Reality (VR) and Augmented Reality (AR), holographic conferencing, virtual education, virtual tourism, smart city, smart home, smart clothes, disaster and emergency management, and work-from-anywhere. This list can be longer with more applications in the upcoming years. Most of the aforementioned applications were originally envisioned for 5G, however, they could not be practically realized. Therefore, it makes sense to address them first while developing the 6G networks.\\

General wireless communications requirements for the given application examples can be defined as: high data rate, high throughput, high capacity, high reliability, low latency, high mobility, high security, low complexity, high connectivity, long battery life, low cost, wide coverage, and more. The importance and priority of the requirements may change under different cases. Moreover, higher levels of performances need to be obtained in next generation systems while meeting the related requirements.\\

Since the requirement diversity is continuously increasing, more sophisticated service types are expected for 6G. Candidate service types are constituted by grouping applications with similar requirements. Examples\footnote{Comprehensive discussions on these potential service types can be found in the given references.} can be given as Big Communications (BigCom), secure uRLLC (SuRLLC), Three-Dimensional Integrated Communications (3D-InteCom), Unconventional Data Communications (UCDC) in \cite{dang2020}; ultra-High-Speed-with-Low-Latency Communications (uHSLLC) in \cite{8760275}; Long-Distance and High-Mobility Communications (LDHMC), Extremely Low-Power Communications (ELPC) in \cite{8766143}; reliable eMBB; Mobile Broadband Reliable Low Latency Communication (MBRLLC), massive URLLC (mURLLC), Human-Centric Services (HCS), Multi-Purpose Services (MPS) in \cite{8869705}. As it is seen from the names, some of the service types (e.g., SuRLLC, uHSLLC, reliable eMMB, MBRLLC, mURLLC, MPS) try to be more inclusive than the 5G service types to serve target applications. It is also possible to see more specific service types (e.g., BigCom, 3D-InteCom, UCDC, LDHMC, ELPC, HCS) in comparison with 5G.\\

The aforementioned applications/services envisioned for 6G illustrates the expected richness of its requirements. These diverse requirements necessitate an ultra-flexible perspective for the incorporation of key enabler technologies and concepts, described below, in future networks.

\section{Ultra-Flexible Perspective for 6G}
\label{sec:s3}

\begin{figure*}[t!]
\centering
\includegraphics[width=17.6cm]{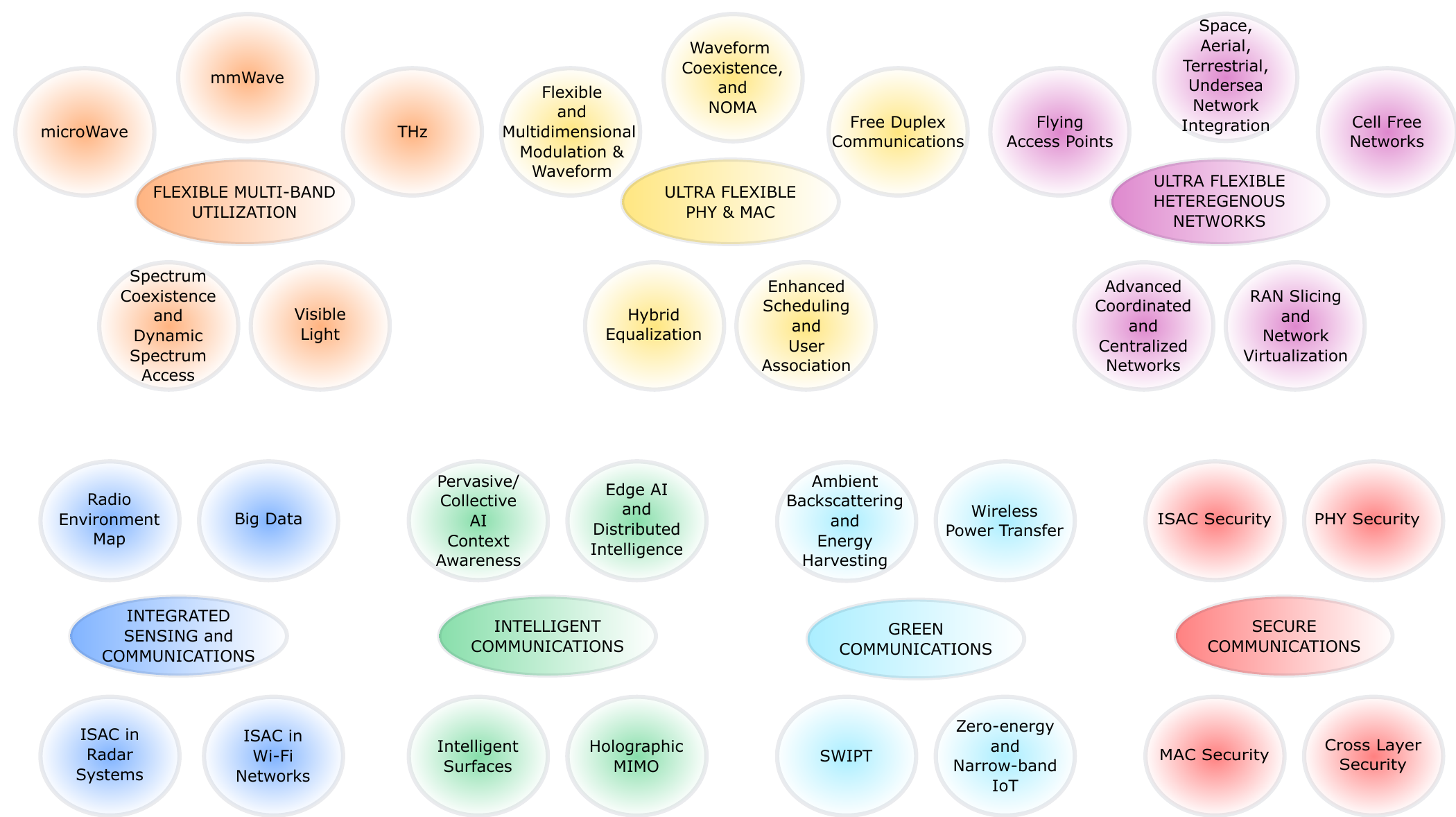}
\caption{Categorization of the promising 6G key enablers under the ultra-flexible perspective.}
\label{fig:s3a}
\end{figure*}

In this section, an inclusive categorization of promising key enablers is presented for 6G communications and their example flexibility aspects are discussed in detail. Then, several flexibility challenges are provided for 6G. Key enabler categories and their related subcategories are shown in Fig.~\ref{fig:s3a}. Many of these technologies are either superficially treated or not studied during 5G standardizations, such as Integrated Sensing and Communications (ISAC) and intelligent communications. Although technologies placed in different categories can have overlapped regions, the given categorization differentiates these technologies regarding their flexibility aspects.\\

Table~\ref{fig:s3b} provides a summary of the example flexibility options achieved by the different technologies. It is worthy to emphasize that the different key enablers have their own impact on the overall flexibility of the system. Ultimately all of them combine together to provide the complete infrastructure capable of realizing the flexible 6G vision that we aspire to achieve.

\subsection{Flexible Multi-Band Utilization}
\label{sec:s3a}

The inclination of communications technologies towards high-frequency bands becomes more appealing due to the increased system capacity and throughput demands of cellular users. Furthermore, flexible usage of available frequency bands, depending on the user and service requirements, is envisioned to be an inherent characteristic of future wireless networks \cite{9083851}.

The millimeter Wave (mmWave) spectrum is starting to be exploited in 5G. It provides new benefits, such as multi-gigabit data rates and reduced interference, however, the use of mmWave bands in 5G is limited by the current International Mobile Telecommunications (IMT) regulations. In World Radiocommunication Conference 2019 (WRC-19), additional 17.25 GHz of spectrum is identified for IMT, where only 1.9 GHz of bandwidth was available before \cite{wrc19}. Therefore, it is expected that spectrum availability in these bands and consequently its flexible utilization will increase during the upcoming years \cite{8798636}. Moreover, beyond 52.6 GHz communications is one of the agenda items for 3GPP Release 17 \cite{8826541}.\\

Frequency bands from 100 GHz to 3 THz are envisioned as a candidate spectrum for 6G communications \cite{8732419}. If THz communications is employed in 6G, it promises a way of dealing with the spectrum scarcity issue by providing an additional degree of flexibility in assigning the most suitable frequency resources for given scenarios \cite{9049790}.\\

Apart from mmWave and THz communications, Visible Light Communications (VLC) also provides spectrum flexibility as a candidate key enabler for 6G networks \cite{9210383, 9083805, 9208801}. Moreover, a new degree of freedom that is information source flexibility is exploited using visible light sources.\\

Spectrum coexistence is another important issue in need of flexible spectrum utilization \cite{9083786, 9083851}. Indeed, the coexistence of cellular communications, Wi-Fi, satellite networks, and radar systems is inevitable in the future  due to both scarce resources and increasing growth in user demands. To exemplify, the coexistence of radar and cellular communications in mmWave frequency bands becomes more popular nowadays \cite{8917703}. Moreover, the idea of Dynamic Spectrum Access (DSA) relies on the spectrum coexistence \cite{9193934}.\\

As it is seen, there are several aspects of flexible multi-band utilization in 6G systems. Flexibility sources can be summarized under three main perspectives: 1) multi-band flexibility, 2) information source flexibility, and 3) spectrum coexistence flexibility.

\subsection{Ultra-Flexible PHY and MAC}
\label{sec:s3b}

One of the unique features of 5G, specifically in the context of PHY design, is the introduction of numerology concept where different configurations of the time-frequency lattice are used to address the varying requirements \cite{yazar2018}. While the numerology concept paves the way for flexibility in beyond 5G networks, it is rather limited considering the competing nature of requirements expected for future 6G networks \cite{9086145}. In addition to the standardized activities, the use of flexible Cyclic Prefix (CP) configurations (e.g., individual CP, common CP, etc.) is explored to enhance the multi-numerology systems for 6G \cite{8861343}.\\

Taking one step beyond the use of different realizations of the same parent waveform as in 5G, multiple waveforms can be accommodated in a single frame for achieving 6G goals \cite{7883931, 9083780}. In line with this, multi-numerology structures can be designed for promising alternative waveforms, that are more suitable for providing additional parameterization options. Having these options enhances flexibility in the PHY layer via increased adaptation capability for meeting a large number of requirements. Moreover, waveform coexistence in the same frame gives the opportunity to serve multiple networks such as radar sensing \cite{mert2020} and Wi-Fi communications together with 6G communications in a flexible manner. There are also several waveform-domain NOMA studies that exploit different resource utilization aspects in the literature \cite{2018-8514314, 9027834, 9145077, 8303689}. Moreover, partial and full overlapping through available resources can also be employed while designing new generation NOMA techniques \cite{8760269, 7088647}. The waveform-domain NOMA concept provides an important flexibility by increasing the resource allocation possibilities in 6G networks \cite{S2352864819304274}. Another flexibility aspect that can arise with 6G is the use of an alternate waveform domain rather than the conventional time-frequency lattice employed by 5G and older generations.\\

In addition to the waveform itself, there is a large number of new generation modulation options in the literature \cite{8631007} and only a small set of them have appeared in the 5G standards. 6G can be enriched with the flexibility provided by these options, particularly Index Modulation (IM) based solutions \cite{dang2020}. This concept can even be extended to multiple domains to provide an additional degree of freedom \cite{8410878}. Moreover, modulation techniques are adaptively designed considering the other key enablers such as Non-Orthogonal Multiple Access (NOMA) \cite{8703780} and Reconfigurable Intelligent Surface (RIS) \cite{8981888} for 6G.\\

Since the configuration of the PHY parameters is, to a large extent, controlled by the Medium Access Control (MAC) layer, it is imperative to develop the flexibility and adaptation capabilities of both layers simultaneously. Two important issues that require flexibility in PHY and MAC would be the ``waveform parameter assignment" or ``numerology scheduling" paradigm under the context of 5G multi-numerology systems \cite{9086145, 8263598}, where the MAC layer is responsible for assignment of parameters of the PHY signal. Similarly, adaptive guard utilization methods have been developed for the MAC layer \cite{8986659, 8946735, yazar2019} to control the new type of interferences in 5G systems. On this basis, it is expected that highly intelligent UE capabilities, and configurable network parameters, and flexible and efficient MAC designs will play a key role in 6G networks due to the expected increased diversity in service types and consequently requirements.\\

Example flexibility perspectives for ultra-flexible PHY and MAC technologies of potential 6G key enablers are given in Table~\ref{fig:s3b}.

\begin{table*}[ht!]
  \begin{center}
    \begin{tabular}{c}
      \raisebox{-\totalheight}{\includegraphics[width=17.6cm]{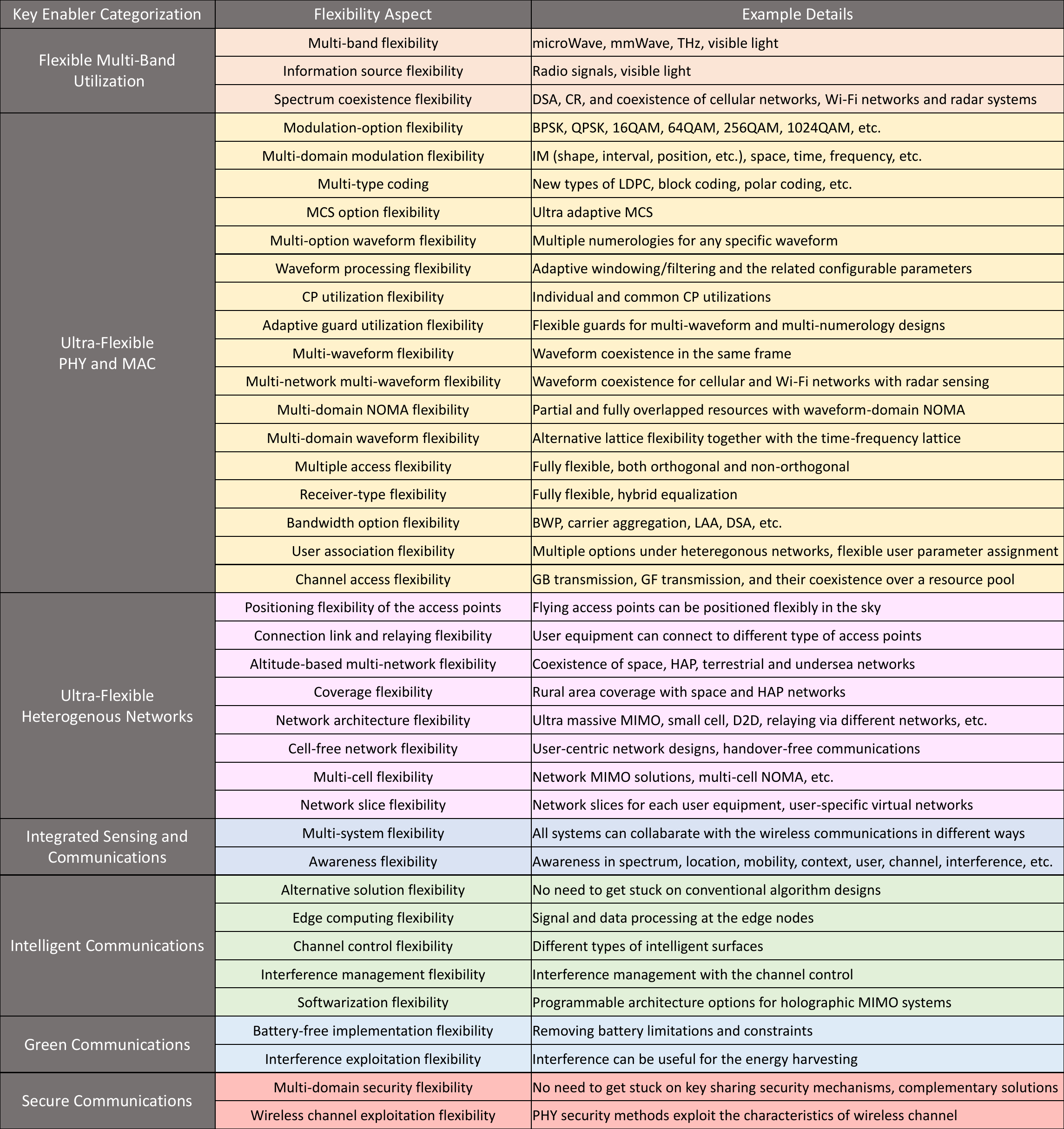}}
    \end{tabular}
    \caption{Example flexibility aspects for the key enabler categories.}
    \label{fig:s3b}
  \end{center}
\end{table*}

\subsection{Ultra-Flexible Heterogeneous Networks}
\label{sec:s3c}

Flying Access Points (FAPs) provide enhanced flexibility for network deployment by allowing dynamic (3-D) positioning of the nodes or even optimized trajectory planning for different objective functions \cite{9055054, 9183763, 9205314}. The push in this direction occurred around the turn of the century \cite{846086}, and was further empowered by projects, such as: 1) Google Loon project, 2) Facebook Aquila project, 3) ABSOLUTE project, 4) Matternet project, and 5) Thales Stratobus project. The integration of FAPs with the terrestrial network can be leveraged to provide coverage in disaster/emergency scenarios, connectivity for rural/isolated areas and capacity enhancement for temporarily crowded places (such as stadiums/concert venues) \cite{9200631}. FAP-based networks are expected to be an important part of 6G not only for achieving deployment flexibility but also for having better wireless propagation provided by a high probability of Line of Sight (LOS) communications \cite{9221119, 8760401}.\\

In addition to the aerial and terrestrial networks, the integration of space (satellite) networks is another aspect of the flexible heterogeneous networks \cite{9174846}. Space networks are also a promising solution for rural area communications \cite{9042251}. They are employed for wireless backhaul communications in the previous cellular networks. However, space networks can also serve aerial user equipment such as drones and UAVs to increase coverage flexibility in 6G systems \cite{9247451}. Moreover, undersea network integration with the other networks will be useful while serving naval platforms.\\

Although, the integration of different networks is ensured, the cell structures of these networks are changing. Cell-less or cell-free networks are one of the potential 6G concepts considering the network architecture richness \cite{7876956,8246848}. User equipment connects to the network via multiple small cells in the cell-less networks. Cell-centric design is transformed into the user-centric system. Hence, it provides both handover-free communications and zero inter-cell interference. Cell-less networks may exploit a new dimension of network Multi-Input Multi-Output (MIMO) flexibility in 6G. As another network MIMO example, advanced coordinated and centralized networks \cite{sohaib2020} are addressed together with NOMA schemes for 6G communications \cite{8352643,8946882}. These networks are called multi-cell NOMA. Flexibility comes with the number of the cells and architecture richness while exploiting other dimensions with NOMA.\\

From the network virtualization perspective, network slices are used in 5G to customize and optimize the network for service types or any other requirement sets \cite{8382171,8676260,8759041}. Hence, the overall performance is increased by meeting different requirement sets with virtually privatized networks. Network slicing brings an important flexibility in 5G since it enables different network options under the same umbrella. The number of network slices can increase for 6G and there may be network slices for each user equipment. This user-centric network slicing architecture can provide full flexibility in the network layer.\\

The number of examples for the flexibility aspects of promising 6G heterogeneous networks can be increased with particular technologies and concepts such as blockchain systems \cite{9184022, 9083784} and quantum communications \cite{8681450} in the future.

\subsection{Integrated Sensing and Communications}
\label{sec:s3d}

\begin{figure*}
\centering
\includegraphics[width=17.6cm]{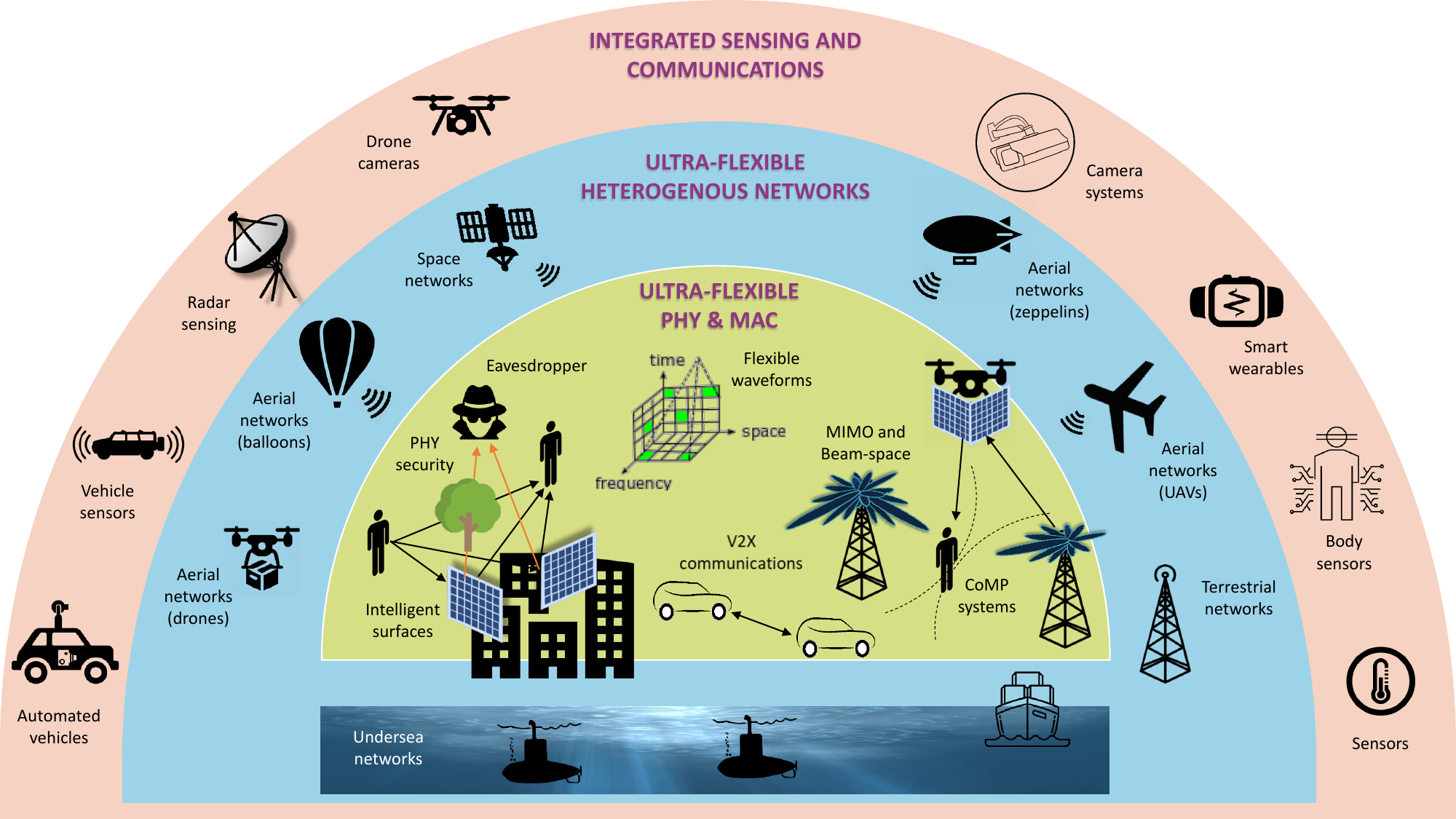}
\caption{The integration of many different sensor hardware with the heterogeneous communications networks under 6G systems.}
\label{fig:s3c}
\end{figure*}

With the emphasis on use cases such as V2X communications in recent years, sensing has attained increased importance leading to the integration of these two applications \cite{9162963}. However, the use of sensing is not limited to V2X or autonomous driving. Rather, if there is any observable data that can be utilized for the optimization or enhancement of the communications systems, it should be leveraged in 6G \cite{halise2020}. The information pertaining to the radio environment can be utilized in improving network deployment, optimizing user association, providing secure communications and so on. Hence, one of the unique novelties of 6G systems is the integration of many different sensor hardware with the heterogeneous communications networks as exemplified in Fig.~\ref{fig:s3c}.\\

While it might sound like a novel idea to some, Integrated Sensing and Communications (ISAC) has been studied in different domains in the past. Cognitive Radio (CR) applications triggered the ISAC research on the last two decades. Spectrum sensing and awareness is one of the first application areas in the ISAC research \cite{4657330}. Location awareness is exploited to improve the wireless communications system design in \cite{4427241}. Satellite and drone images can be used to predict channel parameters \cite{8772043}. Context-awareness is used to optimize network architectures in wireless communications \cite{7980669}. ISAC systems are studied for radar sensing \cite{mert2020,8972666} and Wi-Fi network coexistence \cite{2006.16534} in the literature. However, the complete list of sensing information that can be useful for the next generation cellular communications systems from the ISAC perspective has not yet been comprehensively studied \cite{halise2020}.\\

A Radio Environment Map (REM) is a realization of the ISAC concept \cite{6685772}. It is mainly used to obtain environmental information in the literature, however, for the next generation systems the REM concept will be generalized from environmental-awareness to complete-awareness. REM may include all sensing information in a multi-dimensional manner for wireless communications networks. To exemplify, REM can be a specialized database for the ISAC. Therefore, the flexibility level of the ISAC systems can be determined by the dimensions in REMs. Each dimension in a REM increase the awareness, allowing better resource utilization. Moreover, control of the configurable options and parameters in different communications layers of 6G can be enhanced by more granular REM information.\\

The complete information and awareness of the environment comes at the cost of a high volume of data, variety of sources and significant processing \cite{S2352864820302418, S2352864820301802}. This necessitates the use of big-data processing techniques \cite{6963801}. A significant challenge, however, in this regard is the overhead of data exchange between the sensing and processing nodes. A centralized solution might not be suitable in such scenarios, rendering the use of edge-computing imperative, particularly for low-latency use cases. Moreover, the usage of Artificial Intelligence (AI) solutions can be helpful while processing big-data at the edge nodes.

\subsection{Intelligent Communications}
\label{sec:s3e}

The usage of AI in the communications society has increased in recent years. Several survey and tutorial papers are published on the usage of Machine Learning (ML) for wireless communications \cite{8382166, 8666641, 8714026, 8743390, 8755300, 8957702, 8976180, 9146540}. AI-aided design and optimization has even been leveraged for the flexible implementation options provided in 5G \cite{9086145}. In many of the studies, AI is put at the center of 6G visions \cite{8808168, 8820755, 8926369, 9023459, 9040202, 9122618, 8941882, 9083770, 9237460, S235286482030242X} to complement the classical methods. Indeed, the use of AI is inevitable to incorporate intelligence in the future networks \cite{7792374, 7886994, 8403948}. AI-aided methods can propose fast and efficient solutions in case enough data is available.\\

AI and ML also find a range of applications in ISAC and REM paradigms to extract information regarding the environment from sensed data. A flexible communications system needs to benefit from the advantages of popular ML approaches such as reinforcement learning, deep learning, and edge computing \cite{9206115, 9061001, 9241414, 9247527}. Especially distributed intelligence (edge AI) with edge computing is a promising paradigm for 6G communications \cite{9205981}. The management of multi-band utilization, MAC layer control, heterogeneous and cell-less networks, and the ISAC systems cannot be done in an all centralized manner. Edge computing will play an important role at that point with the help of distributed intelligence so 6G big data can be processed at the edge nodes without being collected at a centralized network.\\

Intelligent networks are not limited to AI-aided concepts. For example, RIS technology is one of the most popular research topics nowadays \cite{8796365,9140329}. Intelligent surfaces bring a new flexibility on the control of channel parameters \cite{9201361}. In the past, a wireless channel was just an observable medium. However, it can be controlled at some level with new generation wireless systems. Interference management flexibility is increased by controlling capabilities of the wireless channel. These flexibility aspects also affect the technology designs in different communications layers \cite{9229054, 9247315}. To exemplify, having a control capability in multipath propagation, such as controlling delay spread, Doppler spread and the number of multipath alleviates the constraints related to waveform design. RIS technology can also be considered as passive holographic MIMO surfaces if it is located closer to the transmitter and receiver antennas \cite{9136592}. Additionally, it is possible to employ holographic MIMO surfaces as active elements. The active holographic MIMO surfaces work similar to massive MIMO but their softwarization flexibility is higher than the conventional MIMO systems \cite{9136592}.

\subsection{Green Communications}
\label{sec:s3f}

While candidate 6G key enablers are increasing the flexibility in different domains, new architectural changes of 6G should support energy efficiency and green communications \cite{8922617, 9222142, 9247128}. Zero-energy Internet of Things (IoT) is one of the most important concepts since ultra low-power wireless communications is necessary for 6G connectivity \cite{9083794}. In this context, Radio Frequency (RF) energy harvesting is studied with ambient backscatter technology for 6G communications \cite{8454398, 8368232}. Thus, low-power wireless systems can obtain their energy from the available high-power radio waves. Backscatter communications enables energy harvesting, simplifying the implementation of zero-energy IoT designs. Provision of rich options for energy-efficiency promises fulfilment of energy requirement variations belonging to different applications. Within this direction, the Symbiotic Radio (SR) concept offers highly reliable backscattering communications together with mutualism spectrum sharing \cite{8907447, 9120455}.\\

It is also possible to benefit from Wireless Power Transfer (WPT) while designing zero-energy IoT systems \cite{7446253}. Under the WPT concept, Simultaneous Wireless Information and Power Transfer (SWIPT) is the most popular technology that may be a candidate for 6G networks \cite{8114544, 9205906}. SWIPT designs are also used for interference exploitation purposes \cite{8214104} since interference can be useful for energy harvesting. Transformation of interference into an energy source introduces another flexibility perspective.

\subsection{Secure Communications}
\label{sec:s3g}

With applications such as eHealth, online banking, and autonomous driving etc., wireless communications promises to be an enabler of innumerable sensitive applications utilizing private data. However, the broadcast nature of wireless communications makes it vulnerable to several security threats such as eavesdropping, impersonation, and jamming. In order to ensure security of such applications, PHY Layer Security (PLS) is an emerging solution that has the capability to complement the conventional cryptography-based security techniques. In fact, PLS is more suited for the increased heterogeneity and power/processing restrictions of future wireless networks since it exploits the characteristics of the wireless channel and PHY properties associated with the link such as noise, fading, interference, and diversity \cite{8509094}. It is also possible to increase this flexibility by designing cross-layer security algorithms with the PHY and MAC layer \cite{7564987}. In several 6G papers, secure communications is discussed as one of the main topics \cite{9023459, 8926369, 9151546, S2352864820302431}. PHY and cross-layer security concepts are expected to play a critical role in 6G networks because of their capability to support joint design of security, reliability, and latency.\\

As discussed in the previous subsections, ISAC and REM concepts will be important enablers in 6G communications. However, a new security problem arises since there may be a large amount of confidential data for ISAC and REM concepts. In the literature, this problem is treated in \cite{9083884} for ISAC security, and in \cite{9102392} for REM security. Thus, there is a need for more secure communications options in 6G networks to meet new types of security requirements, especially for ISAC and REM concepts. Moreover, in order to tackle spoofing attacks, authentication at the physical layer by using features of channel and hardware impairments can also provide a fast, lightweight, and efficient alternative for crypto-security for authentication in future wireless networks. Furthermore, the physical layer solution will also provide efficient robustness against jamming attacks using terrestrial and flying relay and other new multi-antenna-based solution.

\begin{figure*}[t!]
\centering
\includegraphics[width=17.6cm]{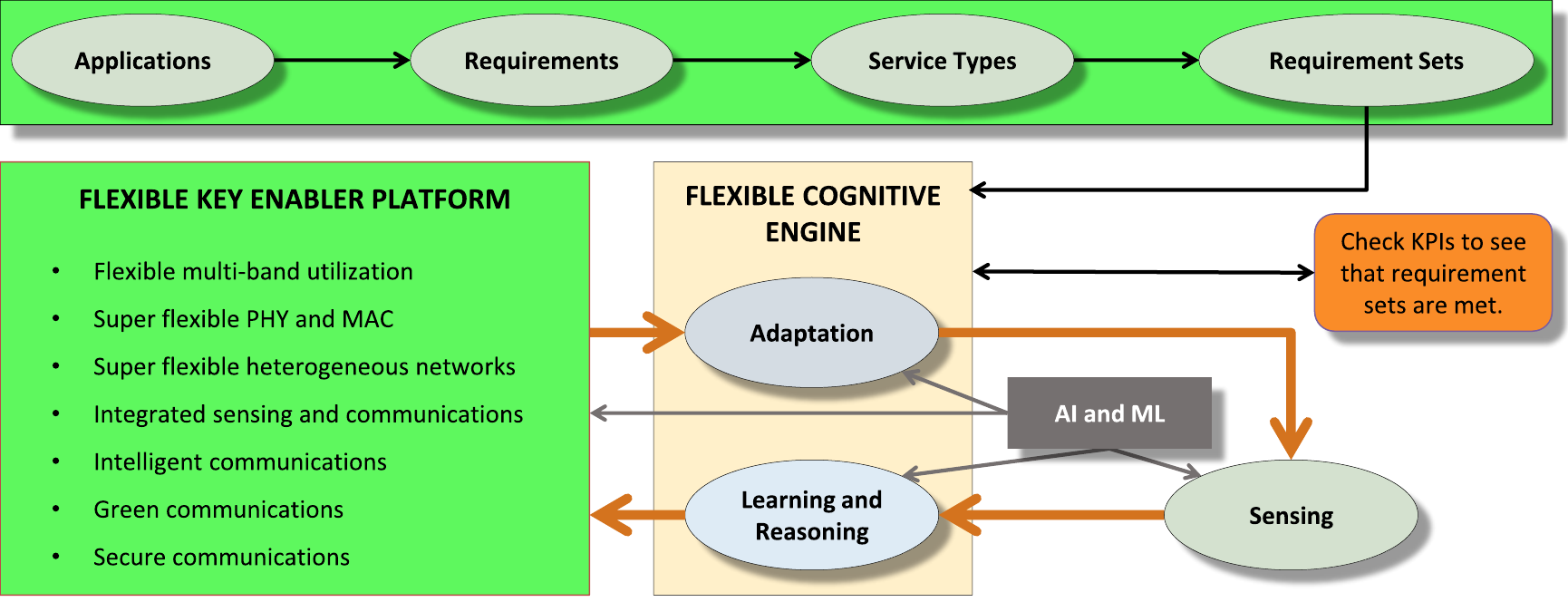}
\caption{The proposed framework includes flexible key enabler platform and flexible cognitive engine. The flexible cognitive engine can be defined as a bridge between the requirements and potential technology options with the related configurations.}
\label{fig:s4a}
\end{figure*}

\subsection{Flexibility Challenges and Opportunities in 6G}
\label{sec:s3h}

The exemplified key enablers show that 6G will have many different flexibility options while 5G systems have limited flexibilities. However, each flexibility is coming with unique challenges. In other words, flexibility opportunities bring new challenges for the 6G networks.

For flexible multi-band utilization, operating the cellular system at multiple frequency bands needs advanced front-end hardware. Additionally, spectrum coexistence of different networks causes new interference problems. If the flexibility challenges on the PHY and MAC layer are investigated, one of the most important problems is the necessity of a flexible waveform system. At that point, either a single but an ultra-flexible waveform can be designed or multiple waveforms can be employed in the same frame. Designing a single waveform to meet all types of requirements did not work for 5G networks. It will be more difficult for 6G with more types of requirements. Moreover, waveform coexistence in the same frame causes new interferences (like inter-numerology interference in 5G). Similarly, partial and fully overlapped NOMA systems have the same interference problem. Control and mitigation of these interferences is expected as another challenge.\\

Flexibility challenges for heterogeneous networks can be exemplified with the developing optimal positioning and relaying algorithms for flying access points. In addition to these algorithms, interference management during the coexistence of different networks is necessary. As another challenge, network MIMO structures provide multi-cell flexibility, however, large amounts of data need to be transferred at the backhaul systems and the amount of burden increases.

As discussed in the previous subsections, ISAC systems can include multiple systems together with the communications networks. Generally, the amount of sensing information increases in parallel to awareness capabilities. However, processing the sensing information causes computational burdens. Additionally, investigating the ways of exploiting this information to enrich the communications systems is another important challenge.

For the flexibility challenges of intelligent communications, first of all, an efficient work distribution between conventional and ML methods is required. A large data sets and useful features need to be developed to make ML mechanisms more functional. Additionally, edge computing algorithm structures should be designed to reduce the workload at transmission points.\\

If we summarize the challenges and opportunities, the following items can be listed:

\begin{itemize}
  \item Need for a rich set of algorithms and techniques at different layers of the protocol stack that are optimized for different applications with their own requirements.
  \item Integration of these rich sets of algorithms into the flexibility framework with minimal overhead and complexity.
  \item Development of techniques that allows flexibility with a simple parameter change without significantly impacting the rest of the system design.
  \item Integration of AI and ML techniques to solve complex system problems together with the classical model based approaches. AI/ML can be applied in different parts of our proposed framework, i.e. it can be applied for better sensing and learning, or for optimal use of the given set of algorithms and approaches, or developing better solutions in the transmission, reception, and modeling of the system.
\end{itemize}

Therefore, there is a need for general frameworks and mechanisms to ease dealing with these challenges all together. Within this direction, an example framework is proposed in the next section.

\section{Ultra-Flexible 6G Framework}
\label{sec:s4}

Gathering together all potential 6G enablers in a flexible framework is an important challenge. Therefore, this section brings the above-mentioned flexible perspectives for the key enabler technologies and concepts under the umbrella of a single ultra-flexible framework for 6G. Here, it is important to realize that the presence of flexible options in itself is not enough to render a network intelligent. Rather, it needs the capability to make best use of the available options. Therefore, some sort of intelligence or cognition is imperative in future wireless networks. Keeping this in mind, the proposed framework has the following primary components: 1) Flexible key enabler platform (like an advanced Mitola radio), 2) flexible cognitive engine, and 3) flexibility performance indicators. Fig.~\ref{fig:s4a} illustrates how these different components are interconnected within the framework. The key points of this framework can be summarized as follows:

\begin{enumerate}
  \item New technologies should be integrated into communications standards via a \textbf{flexible key enabler platform} without waiting for ten years. 
  \item Key enabler technologies should work together in an optimal flexibility to meet different requirements. Therefore, a \textbf{flexible cognitive engine} can make an optimization between different flexibility aspects.
  \item The amount of flexibility needs to be measured while making an optimization. Hence, developing new \textbf{flexibility performance indicators} is necessary.
\end{enumerate}

The previous cellular communications generations were standardized approximately ten years apart. From a different point of view, it took about a decade for the available technologies to be included in the cellular standards. Waiting up to ten years to benefit from an available technology does not make sense if it is possible to develop a platform that hosts different technologies flexibly. For now, we need to tolerate the limited flexibility of 5G technologies for the next decade. However, an advanced Mitola radio can work like a smart phone that has installable and updateable software. We call this radio a \textbf{flexible key enabler platform}. In this concept, the platform has the ability to have new key enabler technologies by a softwarization. Thus, the flexibility level of the wireless communications system can be enhanced with new technologies and the related updates.\\

As it is shown in Fig.~\ref{fig:s4a}, each technology can bring different perspectives to the overall flexibility. There is a need for a multi-objective optimization unit to control all configurable and flexible aspects of the enablers in the flexible key enabler platform. This engine can be designed in an AI-aided manner to optimize the key enabler flexibilities jointly. An optimum work distribution should be done for the flexible configurations of key enablers to meet all the system requirements in the most efficient way. At the end, all system requirements should be met optimally. The \textbf{flexible cognitive engine} will guarantee this optimization by the help of Key Performance Indicators (KPIs) that show the success while meeting requirements. This flexibility optimizer considers also complexity requirements while operating the system.\\

ISAC technologies will be an important part of 6G technologies as discussed in the previous section. Any sensing information can be exploited to make the wireless communications more effective. The flexible cognitive engine can give decisions with more available information while meeting different requirements and handling with several impairments and constraints. Sensing information increases the awareness and controlling capabilities of the system. To provide these capabilities, AI tools in the flexible cognitive engine provide useful and unnoticeable relationships without heuristic designs and theoretical analysis. Hence, the flexible cognitive engine needs three important elements while optimizing the flexibility level with key enablers: 1) Sensing information to increase awareness and controlling capabilities, 2) AI tools to increase the functionality and effectiveness of sensing information, and 3) KPIs to monitor the overall system.

KPIs are needed to measure several performances of the communications system. One of these KPIs can be the \textbf{flexibility performance indicator} so that the achieved flexibility can be quantified. It is difficult to decide on a specific flexibility performance indicator because there are many different flexibility perspectives as shown in Table~\ref{fig:s3b}. This indicator can be technology-specific and require separate metrics for different technology categories. 6G networks will need flexibility indicators similar to the other KPIs such as spectral efficiency and reliability. Generally, the current key enabler technologies are not designed to be called flexible technologies. Flexibility aspects of these key enablers are described mostly based on the inferences. In ideal conditions, 6G technologies need to be designed considering the flexibility perspective as one of the key criteria. At that point, flexibility performance indicators should be employed to quantify the advantages and disadvantages of new designs in both the PHY and MAC layer.

\section{Conclusion}
\label{sec:s5}

5G systems were characterized by diverse applications and requirements. 6G is expected to continue in the same vein by enriching the application fabric even further. Fulfilling such a wide variety of use cases is not possible unless flexibility is incorporated in the promising key enabling technologies for the future networks. Driven by this, we have presented example flexibility aspects for  the potential 6G key enablers under a unique categorization.\\

We believe that 6G should be approached with flexibility at its primary design criterion. Flexibility aspects of the potential key enablers need to play a leading role in the design stages of 6G systems. To this end, we have presented a general framework comprising of the aforementioned flexible key enablers, empowered by a flexible cognitive engine and supported by different aspects of sensing and AI. We believe that the presence of flexible options is imperative but only that is not enough to support the future applications. The ability to extract information regarding the operating environment and making related intelligent decisions are  the way forward in the wireless communications realm. The best possible utilization of the flexibility offered by the key enablers is determined with this vision.\\

The realization of a flexible key enabler platform like the one mentioned above is, however, not straightforward. It requires the methods capable of performing efficient multi-objective optimization to address the various competing applications requirements. Furthermore, quantifying the flexibility by proposing novel performance indicators also remains a significant challenge on the way to ensure a fully-functional flexible, cognitive wireless communications network.

\section*{Acknowledgement}
\label{sec:ackn}

The authors would like to thank Muhammad Sohaib J. Solaija for his valuable comments and suggestions to improve the quality of the paper.

\newpage
\section*{Authors}
\label{sec:auth}

\begin{wrapfigure}{l}{0.25\columnwidth}
  \begin{center}
    \includegraphics[width=0.35\columnwidth]{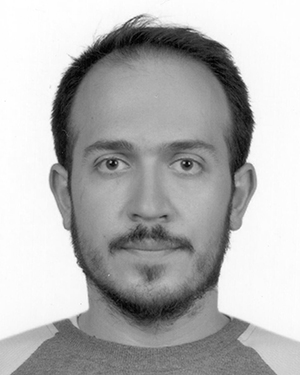}
  \end{center}
\end{wrapfigure}
\textbf{A. Yazar} received his B.Sc. degree in electrical engineering from Eskisehir Osmangazi University, Eskisehir, Turkey in 2011, M.Sc. degree in electrical engineering from Bilkent University, Ankara, Turkey in 2013, and Ph.D. degree in electrical engineering from Istanbul Medipol University, Istanbul, Turkey in 2020. He is currently general coordinator as a member of the Communications, Signal Processing, and Networking Center (CoSiNC) at Istanbul Medipol University. His current research interests are radio resource management techniques and the role of machine learning in wireless communications systems.

\ITUpar

\begin{wrapfigure}{l}{0.25\columnwidth}
  \begin{center}
    \includegraphics[width=0.35\columnwidth]{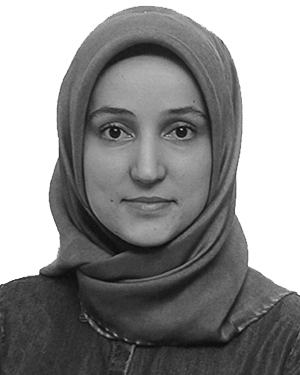}
  \end{center}
\end{wrapfigure}
\textbf{S. Do\u{g}an Tusha} received the B.Sc. degree in electronics and telecommunication engineering from Kocaeli University, Kocaeli, Turkey, in 2015, and the Ph.D. degree in electrical and electronics engineering from Istanbul Medipol University, Istanbul, Turkey, in 2020. She is currently a post-doctoral researcher in the Communications, Signal Processing, and Networking Center (CoSiNC) at Istanbul Medipol University, Istanbul, Turkey. Her research interests include index modulation, millimeter-wave frequency bands, nonorthogonal multiple accessing (NOMA), and random access techniques for next-generation wireless networks.

\ITUpar

\begin{wrapfigure}{l}{0.25\columnwidth}
  \begin{center}
    \includegraphics[width=0.35\columnwidth]{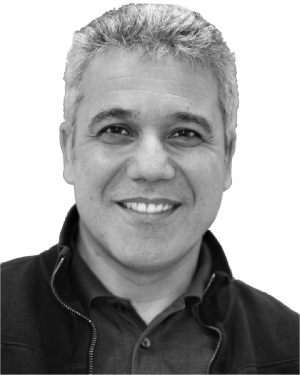}
  \end{center}
\end{wrapfigure}
\textbf{H. Arslan} received the B.S. degree from Middle East Technical University, Ankara, Turkey, in 1992, and the M.S. and Ph.D. degrees from Southern Methodist University, Dallas, TX, USA, in 1994 and 1998, respectively. From 1998 to 2002, he was with the Research Group, Ericsson Inc., NC, USA, where he was involved with several projects related to 2G and 3G wireless communications. Since 2002, he has been with the Electrical Engineering Department, University of South Florida, Tampa, FL, USA. He has also been the Dean of the College of Engineering and Natural Sciences, Istanbul Medipol University, since 2014. He was a part-time Consultant for various companies and institutions, including Anritsu Company, Morgan Hill, CA, USA, and T\"{U}B\.{I}TAK, Turkey. His research interests are in PHY security, mmWave systems, multicarrier communications, co-existence issues on heterogeneous networks, and aeronautical communications.

\end{document}